\documentstyle[epsfig]{mn}
\newif\ifAMStwofonts
\AMStwofontstrue
\def\be{\begin{equation}}
\def\ee{\end{equation}}

\def\gtsima{$\; \buildrel > \over \sim \;$}
\def\ltsima{$\; \buildrel < \over \sim \;$}
\def\prosima{$\; \buildrel \propto \over \sim \;$}
\def\gsim{\lower.5ex\hbox{\gtsima}}
\def\lsim{\lower.5ex\hbox{\ltsima}}
\def\simgt{\lower.5ex\hbox{\gtsima}}
\def\simlt{\lower.5ex\hbox{\ltsima}}
\def\simpr{\lower.5ex\hbox{\prosima}}

\title[The Energy Cascade from Warm Dark Matter Decays]{The Energy Cascade from Warm Dark Matter Decays}         
\author[Vald\'{e}s \& Ferrara]
{M. Vald\'{e}s$^1$ and A. Ferrara$^1$\\
$^1$ SISSA/ISAS, via Beirut 2-4, 34014 Trieste, Italy.}

\begin{document}
\maketitle
\label{firstpage}
\begin{abstract}
We use a set of Monte Carlo simulations to follow the cascade produced by a primary electron of energy $E_{in}$ 
in the intergalactic medium. We choose $E_{in}=3-10$~keV as expected from the decay of one of the 
most popular Warm Dark Matter (WDM) candidates, sterile neutrinos.  Our simulation takes into account 
processes previously neglected such as free-free interactions with ions and recombinations and uses the 
best available cross sections for collisional ionizations and excitations with H and He
and for electron-electron collisions. We precisely derive the fraction of the primary electron energy that 
heats the gas, ionizes atoms and produces line and continuum photons as a function of the ionization fraction.
Handy fitting formulae for all the above energy depositions are provided.
By keeping track of the individual photons we can distinguish between photons in the Ly$\alpha$ resonance
and those with energy $E <$ 10.2 eV that do not interact further with gas.
This separation is important because a Ly$\alpha$ background can heat or cool the gas 
depending on the nature of the photons, and can have effects on the 21~cm radiation emitted 
by neutral H, which will probably become detectable at $z >$ 6 in the near future by the next
generation radio interferometers.  
\end{abstract}


\section{Introduction}

The determination of the gas temperature and ionized fraction in the intergalactic medium (IGM) and interstellar medium (ISM) 
is fundamental for a number of astrophysical studies. In particular, it is essential for the investigation of the nearly uniform, 
dark, neutral state of the Universe known as Dark Ages, which has become in the last decade one of the most 
studied topics in cosmology.
The cosmic phase between hydrogen recombination at $z\sim  $ 1000 and the so-called Epoch of Reionization (EoR) at $z \sim $ 9 can be 
directly
detected by the 21~cm hyperfine triplet-singlet level transition of the ground state of neutral hydrogen. A new generation of 
low frequency radio interferometers such as the Low frequency Array (LOFAR), , the 21 Centimeter Array (21CMA), the Mileura Wide-field 
Array (MWA) and the Square Kilometer Array (SKA), are expected to reach the sensitivity required to map the HI 
distribution at angular resolution of the order of a few arcminutes (e.g. Pen, Wu, \& Peterson 2004; Bowman, Morales, 
\& Hewitt 2005; Kassim et al. 2004; Wyithe, Loeb, \& Barnes 2005). 
If a few issues such as foreground removal, interferences from human-generated signals and ionospheric corrections
will be successfully taken care of these large radio-arrays will be able to perform a tomography of the Universe before
and during the EoR. 

It is therefore crucial to be able to predict the magnitude of the signal. The standard scenario predicts the gas temperature, $T_K$, 
to decouple from the Cosmic Microwave Background (CMB) temperature, $T_{CMB}$, at $z\sim $ 300. $T_K$ then starts decreasing 
adiabatically until the first sources of light heat again the gas above
$T_{CMB}$ at $z \sim $ 20. In this case the 21~cm radiation would be seen in absorption against the CMB in the redshift interval 
20 $<z<$ 300 
as discussed in detail by many recent studies (e.g. Loeb \& Zaldarriaga 2004).

A number of more complex theoretical scenarios have been investigated recently. 
In particular if there was a mechanism to heat the gas at $z\geq $ 10, thus producing a thermal history different from the standard 
one, a direct imprint would be left on the 21~cm radiation from the neutral gas,
which could then be observable in emission rather than in absorption.

First stars (Ciardi \& Salvaterra 2007), intermediate mass black holes (Zaroubi et al. 2006; Ripamonti, private communication) 
and decaying or annihilating Warm and Cold Dark Matter (CDM) particles
(Shchekinov \& Vasiliev 2006; Furlanetto, Oh, \& Pierpaoli 2006;  Valdes et al. 2007) 
could be capable of leaving a significant trace in the evolution of the neutral intergalactic gas.
Future 21~cm observation will possibly rule out or confirm these theoretical predictions, opening a new exciting frontier in cosmology 
and possibly unveiling the nature of dark matter particles.

All the most popular dark matter particle candidates inject energy into the IGM, either via decays or annihilations, initiating an 
energy cascade from energetic primary photons or electrons. The energy deposition depends on the large number of interactions taking 
place during the propagation of the cascade particles through the IGM. It is therefore very important for achieving correct results 
to follow in detail these cascades and to know exactly how much of the energy injected ionizes the gas, produces radiation by 
collisional excitations, and  heats surrounding medium, respectively. 

This fundamental problem has received attention in the past (e.g. Bergeron \& Collin-Souffrin 1973), and the results 
achieved were corrected by an extensive study by Shull and van Steemberg (1985), which we will denote hereafter as SVS85, 
which was a development of a previous work by Shull (1979), hereafter S79.  

In this work we present the results of a Monte Carlo calculation in the spirit of the one from S79 and SVS85; in comparison, we 
use more recent and accurate cross sections for collisional ionization and excitations from electron impacts, for electron-electron 
collisions, for free-free interactions and recombinations. In addition to this we follow in detail the radiation produced by the 
excitations and we are able to predict precisely how much of the energy goes into photons that do not further interact 
with the gas (E$\leq $ 10.2 eV) and how much of it contributes to the Ly${\alpha}$ background, which directly affects
the physics of the 21~cm line radiation by the so called Wouthuysen-Field effect
(e.g. Wouthuysen 1952; Field 1959; Hirata 2005) and which can heat or cool the gas depending 
on the nature of the Ly$\alpha $ photons (Chen \& Miralda Escud$\acute{e}$ 2004, Chuzhoy \& Shapiro 2007).

\section{Method}

\subsection{Description of the calculation}

We calculate the effects produced by an X-photon of $\sim $keV energy injected into the IGM with $T\ll 10^4$ K.
At these energies the dominant interaction is photoionization (see e.g.
Zdziarski \& Svensson 1989) and the X-photon ionizes an H or He atom producing an energetic primary electron. 
We follow the subsequent secondary cascade products.

For our calculations we choose two specific energies for the primary electron, $E_{in}=3$ keV and $10$ keV.
This energy range is of great interest because sterile neutrinos, one of the
most promising WDM candidates, are expected to emit line radiation at an energy
between 3-25 keV. A large effort has been done recently by several authors to constrain the mass and lifetime
of radiatively decaying sterile neutrinos from X-ray observations (Abazajian, Fuller \& Tucker 2001; 
Abazajian 2006; Abazajian \& Koushiappas 2006; Boyarsky et al. 2006b; Watson et al. 2006; Mapelli \& Ferrara 2005; 
Boyarsky et al. 2006a). 

Once the primary electron is injected into the IGM the code calculates the cross sections relative to a list
of possible processes: H, He, HeI ionization;  H, He excitation; collisions with thermal electrons; free-free interactions with
ionized atoms; recombinations. We then estimate the mean free paths and derive the probability for a single 
electron to have any of the aforementioned interactions.
We assume, in agreement with the 3-yr Wilkinson Microwave Anisotropy Probe (WMAP) data analysis, 
that the helium fraction by mass is $f_{He}=$ 0.248 (Spergel et al. 2007)
and we follow the fate of a primary electron and its secondary
products for different values of the ionized fraction $x_e$ of the gas, which, as showed by SVS85 and S79, is the free parameter
which affects the results.
We assume, similarly, that $x_e \equiv n\mbox{(H}^+\mbox{)}/n\mbox{(H)}\equiv n\mbox{(He}^+\mbox{)}/n\mbox{(He).}$
   
For each primary energy $E_{in}$ and for each assumed value of the gas ionized fraction $x_e$ we 
performed 1000 Monte Carlo realizations, a number 
sufficient to produce consistent results not biased by the random nature of the computation. We will return on this aspect
at the end of this Sec.

If the primary electron (or an energetic secondary electron) collisionally ionizes an H or He atom
then the resulting two electrons have to be followed separately as they interact further with the gas.

Once the electron energy has degraded  below 10.2 eV we assume that its entire energy is deposited as heat. This is a 
simplification as the electron would actually thermalize with the gas and as a consequence the precise heating should 
be calculated by taking into account the gas temperature.
If the temperature is of the order of $10^4$~K electrons with energies lower than 1 eV could even cool the gas.

Collisional excitations of H ad He produce photons that escape freely in the surrounding medium  
if their energy is lower than 10.2 eV or that can further interact with the gas if they have higher energy. 

The study from S79 and SVS85 derives the amount of energy which is deposited in excitations but does not give details
about the individual photons. We want to estimate instead the amount of energy that goes into Ly$\alpha$ photons: this radiation
can in fact interact further with the gas by the Wouthuysen-Field effect and is crucial to correctly
estimate the 21~cm signal. Furthermore, Ly$\alpha$ photons by scattering resonantly off neutral hydrogen
can cool or heat the gas, depending on whether they enter the resonance from its red or blue wing respectively,
as we will explain more in detail later in this Letter.

An additional novel feature of our model is the inclusion of the previously neglected two-photon forbidden 
transition $2s \rightarrow 1s$. We included both
the direct collisional excitation cross section to the $2s$ level and the probability that a collisional excitation to a level 
$n\geq $ 3 results in a cascade through the $2s$ level rather than through $2p$.
As noted recently (Hirata 2005, Chuzhoy \& Shapiro 2007) this effect is not negligible in general and constitutes the most probable 
decay channel for H atoms excited to a level $n=$ 3 or higher. The result is the emission of two 
photons below the Ly$\alpha$ energy that do not further interact with the gas.

\begin{table*}\label{table1}
  \begin{center}
    \begin{tabular}{|c|c|c|c|c|}\hline
      \hline 
       $x_e$             & Gas     &  Excitations     & Ionizations      &      Excitations \\
      (ionized fraction) & Heating & (Lyman-$\alpha$) & (H, He, HeII)    & ($E<10.2$ eV)\\
      \hline 
      0.00010 &	  0.136566$\pm$0.006526	& 0.334972$\pm$0.012904	& 0.379666$\pm$0.013697	& 0.148796$\pm$0.006668 \\
      0.00015	& 0.143090$\pm$0.007234	& 0.330344$\pm$0.012740	& 0.379530$\pm$0.012823	& 0.147036$\pm$0.006585\\ 
      0.00020	& 0.148240$\pm$0.006846 & 0.326795$\pm$0.012202	& 0.379611$\pm$0.013011	& 0.145354$\pm$0.006477\\
	  0.00030	& 0.155762$\pm$0.007440	& 0.322317$\pm$0.012472	& 0.378449$\pm$0.012658	& 0.143472$\pm$0.006310\\
	  0.00050	& 0.166903$\pm$0.007935	& 0.315132$\pm$0.012997	& 0.377684$\pm$0.013238	& 0.140281$\pm$0.006595\\
	  0.00100	& 0.183868$\pm$0.008404	& 0.305810$\pm$0.012733	& 0.374306$\pm$0.012698	& 0.136016$\pm$0.006497\\
	  0.00150	& 0.196387$\pm$0.008972	& 0.298643$\pm$0.012996	& 0.372148$\pm$0.012737	& 0.132822$\pm$0.006566\\
	  0.00200	& 0.204938$\pm$0.009070	& 0.294879$\pm$0.012814	& 0.368847$\pm$0.012509	& 0.131336$\pm$0.006457\\
	  0.00300	& 0.219333$\pm$0.009728	& 0.286883$\pm$0.012733	& 0.366153$\pm$0.012303	& 0.127631$\pm$0.006581\\
	  0.00500	& 0.239705$\pm$0.010887	& 0.277359$\pm$0.013153	& 0.359742$\pm$0.012514	& 0.123194$\pm$0.006770\\
	  0.01000	& 0.271348$\pm$0.012790	& 0.264273$\pm$0.013665	& 0.347151$\pm$0.012160	& 0.117228$\pm$0.006788\\
	  0.01500	& 0.295171$\pm$0.013789	& 0.254455$\pm$0.013040	& 0.337623$\pm$0.011888	& 0.112751$\pm$0.006453\\
	  0.02000	& 0.314125$\pm$0.015099	& 0.246530$\pm$0.013916	& 0.330200$\pm$0.012187	& 0.109145$\pm$0.006836\\
	  0.03000	& 0.343381$\pm$0.016290	& 0.235511$\pm$0.013593	& 0.316941$\pm$0.012406	& 0.104166$\pm$0.006984\\
	  0.05000	& 0.388884$\pm$0.019726	& 0.217985$\pm$0.014083	& 0.296736$\pm$0.013365	& 0.096396$\pm$0.007000\\
	  0.10000	& 0.470896$\pm$0.025642	& 0.188577$\pm$0.014297	& 0.257177$\pm$0.014095	& 0.083350$\pm$0.006941\\
	  0.15000	& 0.535092$\pm$0.027713	& 0.165122$\pm$0.014062	& 0.226825$\pm$0.016066	& 0.072962$\pm$0.006617\\
	  0.20000	& 0.590440$\pm$0.029182	& 0.145668$\pm$0.013954	& 0.199731$\pm$0.016245	& 0.064162$\pm$0.006753\\
	  0.30000	& 0.676904$\pm$0.027731	& 0.114286$\pm$0.012867	& 0.158192$\pm$0.015455	& 0.050618$\pm$0.006023\\
	  0.50000	& 0.807223$\pm$0.022754	& 0.067655$\pm$0.009607	& 0.095154$\pm$0.013304	& 0.029968$\pm$0.004703\\
	  0.70000	& 0.900674$\pm$0.014926	& 0.034983$\pm$0.006577	& 0.048889$\pm$0.009166	& 0.015454$\pm$0.003125\\
	  0.80000	& 0.937668$\pm$0.010710	& 0.022106$\pm$0.004776	& 0.030452$\pm$0.006889	& 0.009774$\pm$0.002281\\
	  0.90000	& 0.970899$\pm$0.006796	& 0.010273$\pm$0.002993	& 0.014284$\pm$0.004777	& 0.004543$\pm$0.001432\\
	  0.95000	& 0.985733$\pm$0.004310	& 0.004954$\pm$0.002056	& 0.007120$\pm$0.003168	& 0.002193$\pm$0.000976\\
	  0.99000	& 0.997217$\pm$0.001808	& 0.000993$\pm$0.000824	& 0.001352$\pm$0.001382	& 0.000439$\pm$0.000424\\
      \hline
    \end{tabular}
  \end{center}
  \caption{Fraction of the energy $E_{in}$ of a 10 keV primary electron that is deposited into heat, ionizations, Ly$\alpha$ line radiation 
and photons with energy $E <$ 10.2 eV after two-photon decay from the hydrogen $2s$ level. For every value of the ionized 
fraction 1000 realizations were performed.} 
\end{table*}

With our code we are therefore able to separate the Ly$\alpha$ radiation from the less energetic photons 
produced by cascade from the $2s$ level.
All Ly$\alpha$ photons that result from collisional excitations  (or \textit{injected} Ly$\alpha$ photons,
see Chen \& Miralda Escud$\acute{e}$ 2004) have a cooling effect on the gas.

We also included in our calculation processes that can produce continuum photons, such as recombinations and Bremsstrahlung 
free-free interactions of electrons with ionized atoms.
This effects are negligible as we will see and there is virtually no production of photons between the Lyman-$\alpha$ and 
the Lyman-$\beta$ resonances. This radiation would have redshifted into \textit{continuum} Ly$\alpha$ photons 
(Chen \& Miralda Escud$\acute{e}$ 2004)
and by entering the Ly$\alpha$ resonance from its blue wing would have heated the gas instead.

Electron-electron collisions between secondary and thermal electrons were implemented as in SVS85 and S79 
and we treated similarly the energy distribution of secondary electrons following collisional ionization of H and He, 
so we refer the reader to those works for a detailed explanation.

As we mentioned earlier, we found that 1000 Monte Carlo realizations gave consistent results. Going from 50 to 1000
realizations changed the averaged values by less than $5\%$ and the respective $\sigma$ by less than $10\%$. We deemed 
1000 to be a sufficient number of realizations for a stable result. The $1 \sigma$ values of the
energy deposition fractions are included in Table 1 and they vary from $2.9 \%$ to $0.04\%$ of the total 
amount of energy.

\subsection{Cross sections}

For our purposes it is important to use the best available cross sections for the several interaction channels considered in our 
calculation.
We will give here the references rather than entering in the description of physical details that are not the scope of this Letter.
The cross sections ($\sigma_i$) for collisional ionization of H, He, He+ were taken from Kim \& Rudd (1994), Shah et al. (1987), 
Shah et al. (1988). A simple functional fit is  

\begin{equation}\label{}
\sigma_i (x)=\frac{4 \pi {a_0}^2}{x}\left[a \ln (x) 
+b \left(1-\frac{1}{x}\right) + c \frac{\ln t}{t+1} \right]
\end{equation}
where $x=E_{kin}/E_B$ is the ratio between the incoming electron energy and the binding energy of the atomic electron.  

The collisional excitation cross sections of H and He are from Kim \& Desclaux (2002), while for the excitation 
to the $2s$ level of H we used the work from Bransden \& Noble (1976).
The cross section for Coulomb collisions between electrons is from Spitzer \& Scott (1969) as in S79, while the free-free
cross section ($\sigma_{ff}$) for electrons interacting with protons is given by the Bethe-Heitler 
quantum-mechanical Born approximated result (see e.g. Haug 1997),

\begin{equation}\label{}
\frac {d \sigma_{ff}}{d k} \approx \frac{16 \alpha Z^2 {r_0}^2}{3 k {p_i}^2} \ln \left(\frac{p_i+p_f}{p_i-p_f}\right) \mbox{,}
\end{equation}
where $m_e$ the electron mass, $r_0$ is the classical electron radius, 
$\alpha$ is the fine structure constant, $k = E/mc^2$  
is the photon energy
in units of $m_e c^2$, and $p_i$, $p_f$ are the momenta of the incident and scattered electron respectively, again in units of 
$m_e c^2$.

\begin{figure*}
  \centerline{\psfig{figure=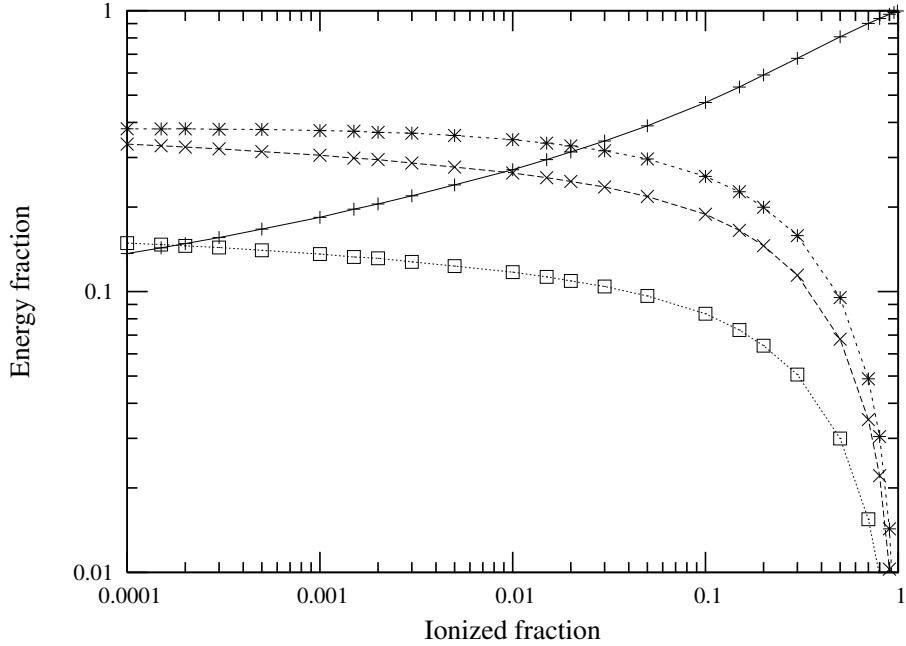,width=12.5cm,angle=0}}
  \caption{Fractional energy losses for a primary 10 keV electron (holds within 2\% also for the 3 keV case) 
going into non-interacting photons with $E <$ 10.2eV (squares), gas heating (pluses), Ly$\alpha$ \textit{injected} photons 
(crosses), and ionizations (asterisks).}
\label{fig:1}
\end{figure*}

As mentioned before we also included the recombination cross section $\sigma_r$, to take into
account a process which could produce continuum photons. Neglecting helium recombinations, we have that for hydrogen  
\begin{equation}\label{}
\sigma_r(\nu,n)\approx 3 \times 10^{10}\frac{g_{fb}}{\nu n^3 v_e^2}\,\, \mbox{cm}^2
\end{equation}
where $g_{fb}$ is the Gaunt factor of $\cal O$(1), $\nu$ is the emitted radiation frequency, $v_e$ is the electron velocity 
and $n$ is the level at which the electron recombines.

\section{Results}

Our results are summarized in Table 1 and in Figure 1. In the Table we report the fraction of the energy of a 10 keV
primary electron which, for different values of the gas ionized fraction $x_e$, 
is deposited into heat, Ly$\alpha$ excitations, ionizations and photons with $E <$ 10.2 eV. The errors correspond the standard 
deviation from the mean calculated over 1000 Monte Carlo realizations of the experiment.

The energy fraction that goes into heating grows rapidly as the gas ionization fraction value becomes higher. The physical 
reason of this behavior is that electron-electron interactions become dominant for high values of $x_e$.

We performed our calculations for 3 keV and 10 keV to account for the theoretically and observationally inferred range of the 
sterile neutrino mass and found that the energy fraction distributed among the different processes remains similar in both 
cases (within $2\%$). This is in agreement with the results from S79 and SVS85 that find that for primary electrons with 
energies higher than 100 eV the fractional energy depositions converge rapidly to a common behavior.

We gave a convenient functional form to our results by fitting the data in Table 1 with an accuracy greater than $3.5$\%. 

\begin{itemize}
\item Fraction $f_h$ of the primary energy deposited into heat:
\begin{equation}\label{}
f_h= 1.0-0.8751\,(\,1.0-x^{0.4052})
\end{equation}
\item Fraction $f_{\alpha}$ of the primary energy converted into Ly$\alpha$ radiation:
\begin{equation}\label{}
f_{\alpha}= 0.3484\,(\,1.0-x^{0.3065})^{0.9533}
\end{equation}
\item Fraction $f_i$ of the primary energy deposited into ionizations:
\begin{equation}\label{}
f_i= 0.3846\,(\,1.0 - x^{0.5420})^{1.1952}
\end{equation}
\item Fraction $f_c$ of the primary energy deposited into continuum radiation:
\begin{equation}\label{}
f_c= 0.1537\,(\,1.0 - x^{0.3224})
\end{equation}
\end{itemize} 

Our results differ sensibly from those of the past studies, with differences as high as 30\% for some 
values of $x_e$. The reason for this is a combination of the better and more modern cross sections which 
we used and of the additional processes included in the calculation such as the transitions through the $2s$ level of H.
We compare in detail our results with those obtained by SVS85 in Fig 2. 
While for small values of $x_e$ the curves are similar it is evident that the differences become substantial 
for $x_e \simgt > 0.1$

\section{Discussion and conclusions}

\begin{figure*}
  \centerline{\psfig{figure=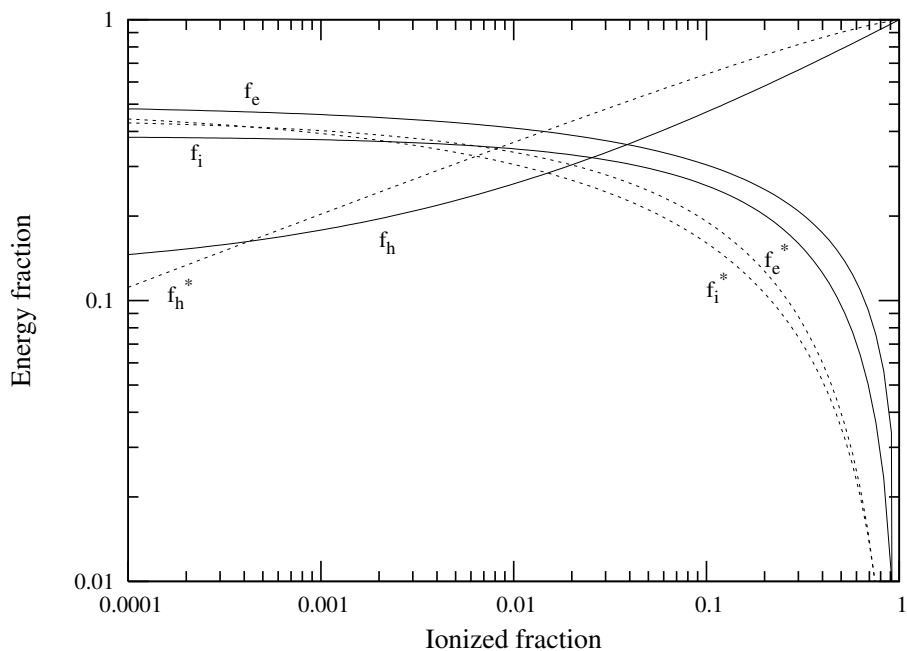,width=12.5cm,angle=0}}
  \caption{Comparison of the present results (solid lines) with SVS85 (dotted) in terms of the fractional energy of 
           the primary electron going into heat, $f_h$, ionizations, $f_i$, and excitations, $f_e$. In the plot we 
	       indicate with $^*$ the fractional energy depositions as fitted in SVS85.}
\label{fig:2}
\end{figure*}

We have presented an updated calculation of the energy cascade arising from a primary electron with energy in the 
range ($3-10$ keV) predicted for one of the most popular dark matter particle candidates, i.e. sterile neutrinos. 
We have computed the fractional energy deposition into ionizations, excitations and heating to a new level of detail 
and followed the fate of individual photons to be able to distinguish between Ly$\alpha$ \textit{injected} radiation 
and continuum photons with energies under 10.2~eV.

As mentioned previously in this work, Ly$\alpha$ radiation affects the gas in several ways. It has a thermal
effect on the matter, with a heating or cooling effect depending on the nature of the photons. 
Line or $injected$ Ly$\alpha$ photons cool the gas, while photons between the Ly$\alpha$ and Ly$\beta$ resonances
that redshift to 10.2~eV (\textit{continuum} Ly$\alpha$ photons) act as a heat source for the gas.

A Ly$\alpha$ background is also responsible for the Wouthuysen-Field process, which directly affects the physics 
of the 21~cm line radiation from neutral H and can make it visible in emission or absorption against the CMB.
This aspect is important because 21~cm observations of the high redshift Universe will be performed in the 
near future by next generation low frequency radio interferometers such as LOFAR.

We included in our calculations the only mechanisms that could produce \textit{continuum} Ly$\alpha$ radiation, 
recombinations and free-free interactions with ions. Both processes are more probable as the ionized fraction 
increases, but at the same time the cross section for electron-electron collisions becomes dominant, so we found 
that both these channels are practically negligible and that electrons injected in a highly ionized gas are 
thermalized before they can produce continuum photons.

We expect that the same calculations performed for relativistic electrons could produce interesting results in this sense, 
also taking into account that processes such as inverse-Compton on CMB photons would become important and produce
continuum radiation. This could be a useful extension to this work and could be applied to study the effects of 
Light Dark Matter decays/annihilations in the energy range around $\sim $ 10~MeV.

\end{document}